\documentclass[]{jfm}
\usepackage{amsfonts}
\usepackage{amssymb}
\usepackage{amsmath,bm}

\usepackage{natbib}

\usepackage{psfrag}
\usepackage{graphicx}
\usepackage{grffile}

\usepackage{hyperref}
\usepackage[final]{pdfpages}
\usepackage{url} 

\usepackage[]{subcaption}
\usepackage{tikz}
\usetikzlibrary{positioning}

\usepackage{authblk}

\definecolor{blue_py}{RGB}{31,119,180}
\definecolor{orange_py}{RGB}{255, 127, 14} 
\definecolor{green_py}{RGB}{44, 160, 44}
\definecolor{red_py}{RGB}{214, 39, 40}

\begin{document}

\title{Viscous lubrication force between spherical bubbles with time-dependent radii}
\author[1]{\small Jean-Lou Pierson}
\affil[1]{\small IFP Energies Nouvelles, Solaize, 69360, France}
\affil[*]{ jean-lou.pierson@ifpen.fr}

\date{\today}

\maketitle 

\begin{abstract}
Motivated by the dynamics of microbubbles in dissolved gas flotation processes, we consider theoretically the approach between two shear-free spherical bubbles with time-dependent radii. We make use of the lubrication assumption to obtain the thin film flow between the bubbles. Our analysis underscores that for the shear-free condition and spherical shape assumption to hold, both the viscosity ratio and the capillary number must be significantly smaller than the thickness of the film. We demonstrate that the lubrication force exhibits weak singular behavior, scaling logarithmically with the ratio of bubble radius to film thickness. To assess the accuracy of our findings, we compare the obtained solution to results from Stokes flow theory. The comparison demonstrates that our current results are reliable, provided that we combine the lubrication forces with subdominant corrections, which require proper matching and computation to the solution far from the film. In practice, we compute these subdominant corrections in the case of two equal bubbles or a bubble close to a plane-free surface either by a curve fit of numerical results from bi-spherical coordinate solutions or by using results from the literature.
We illustrate the relevance of the solution to determine the drainage time of a small bubble rising to a free surface and the drainage rate of expanding bubbles under force-free conditions.  
Finally, in the discussion, we relax the assumption of negligible shear and show that even a small but non-negligible shear induced by fluid motion within the bubble introduces a singular term in the lubrication force.\end{abstract}


\section{Introduction}
\label{sec:intro}

The coalescence of bubbles is ubiquitous in nature and industry. For instance, the coalescence of bubbles at a free surface significantly affects the production of sea aerosol \citep{deike2022}. In the industry, the size of bubbles profoundly impacts the efficiency of flotation processes \citep{nguyen2003}. These flotation processes are particularly interesting for recovering microparticles like microplastics or fine particles \citep{swart2022}. However, one major limitation of this technique is the requirement for generating small bubbles to capture the smallest particles, as emphasized by \citet{yoon1989}. 
Dissolved gas flotation is one existing technology used to generate micron-sized bubbles. In this process, the pressure of a liquid containing dissolved air is reduced, thereby releasing micron-sized bubbles that grow in size as they translate. The dynamics of these bubbles with time-dependent radii, especially when they are nearby and about to coalesce, strongly influence the process efficiency, which motivates the present study. The current investigation also holds relevance in the investigation of coalescing bubbles within magma or the coalescence phenomenon observed in water electrolysis processes. 

The hydrodynamic resistance between two spherical drops or bubbles in the Stokes regime has been studied by \citet{haber1973} using a bispherical coordinate solution of the Stokes equation. \citet{davis1989} used lubrication theory to obtain the leading order forces on normally translating drops when the film thickness is significantly smaller than the drop radius. They observed good agreement between their results and the bispherical solution for sufficiently thin films. They also highlight the effect of the viscosity ratio on the force and mobility of the interface. In particular, the interface can be considered fully mobile or shear-free only for very small viscosity ratios.
 
\citet{chesters1982} investigated the film drainage between two approaching bubbles, assuming the interfaces to be deformable and shear-free. Their study revealed two distinct regimes of film drainage based on whether viscosity or inertia dominates. Coalescence occurs within a finite time, even without considering van der Waals effects, and a dimple is observed in the latter regime. Instead, in viscous-dominated regimes, no dimple is observed. \citet{pigeonneau2011} performed simulations using boundary integral methods to study the rise of a deformable bubble toward a free surface in viscous-dominated flows. Their findings indicated a significant influence of deformation on the dynamics of film drainage compared to the results obtained from the bi-spherical solution. The deformation tended to delay the drainage of the film.

Translating bubbles with time-dependent radii has received relatively less attention in the literature. \citet{michelin2018} derived a solution using bi-spherical coordinates for two bubbles with time-dependent radii. Their analysis revealed that hydrodynamic effects are mostly negligible, except when the bubbles are in close proximity. In the nearly inviscid limit \citet{van2002} derived the force on translating bubbles with time-dependent radii near a plane wall. Through an extended Rayleigh-Plesset equation, they investigated the trajectory of a bubble with an oscillating radius.

Focusing on translating spherical bubbles with time-dependent radii and shear-free interfaces in the Stokes regime, the present study aims to disentangle the various contributions to the forces. For this purpose, we use lubrication theory to derive closed-form solutions to the forces. In addition, we validate our theory by comparing it with Stokes flow theory. We then apply our results to two canonical configurations from existing literature, specifically examining the scenarios of a bubble approaching a free interface and the interaction of two growing bubbles under force-free conditions. The problem and current assumptions are introduced in section \ref{sec:problem}. 
In section \ref{sec:lubrication} we consider the asymptotic limit of a thin film.   
Then, the forces on the bubbles are derived and compared to Stokes flow solutions in section \ref{sec:force}. Predictions from the theory in two different physical configurations are discussed in section \ref{sec:examples}. The outcomes and assumptions of our investigation are discussed in section \ref{sec:conc}. Specifically, we relax the assumption of shear-free interfaces by considering small but non-negligible shear at the bubbles interfaces.

\section{Formulation of the problem}
\label{sec:problem}

\subsection{Description of the problem}

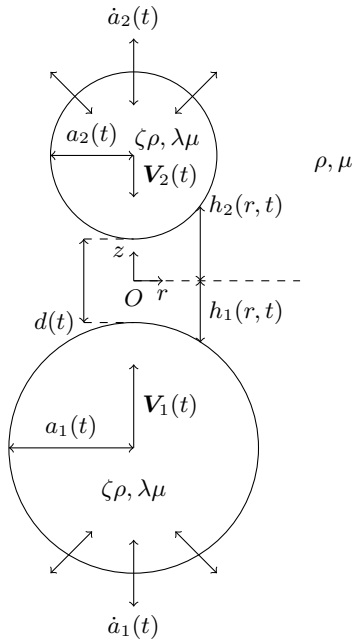
\begin{figure}
\centering
\begin{tikzpicture}[scale=1.1]

\draw[<->] (0.5,2) -- (1,2.5);
\draw[<->] (0,2.1) -- (0,2.9) node[above] {$\dot{a}_2(t)$};
\draw[<->] (-0.5,2) -- (-1,2.5) ;

\draw [] (0,1.5) circle (1);
\draw[<->] (0,1.5) -- (-1,1.5) node[midway,above] {$a_2(t)$};
\draw[->] (0,1.5) -- node [right] {$\bm{V}_2(t)$} (0,1) ;

\draw[<->] (0.5,-3) -- (1,-3.5);
\draw[<->] (0,-3.1) -- (0,-3.9) node[below] {$\dot{a}_1(t)$};
\draw[<->] (-0.5,-3) -- (-1,-3.5) ;

\node[] at (2.4,1.4) {$\rho,\mu$};
\node[] at (0.4,1.7) {$\zeta\rho,\lambda\mu $};
\node[] at (0,-2.5) {$\zeta\rho,\lambda\mu$};

\draw [] (0,-2) circle (1.5);
\draw[<->] (0,-2) -- (-1.5,-2) node[midway,above] {$a_1(t)$};
\draw[->] (0,-2) -- node [right] {$\bm{V}_1(t)$} (0,-1) ;




\draw[<->] (0.8,0)  -- (0.8,0.89) node[right] {$h_2(r,t)$};


\draw[<->] (0.8,-0.74) -- node[right] {$h_1(r,t)$} (0.8,0) ;

\draw[<->] (-0.6,-0.5)node[left] {$d(t)$} -- (-0.6,0.5) ;
\draw[dashed] (-0.6,-0.5) -- (0,-0.5);
\draw[dashed] (-0.6,0.5) -- (0,0.5);

\draw[->] (0,0) node [below] {$O$} --(0,0.35) node [left] {$z$};
\draw[->] (0,0) --(0.35,0) node [below] {$r$};

\draw[dashed] (0,0)--(2,0);


\end{tikzpicture}
\caption{Two bubbles with radii $a_1$ and $a_2$ in relative motion.}
\label{fig:scheme}
\end{figure}

We consider the thin film flow between two translating spherical bubbles moving along their line of centers (Figure \ref{fig:scheme}). The radius of the bubble $i$ ($i=1,2$) is denoted $a_i(t)$, and the rate of change of the bubble radius is noted $\dot{a}_i(t)$. 
The bubbles center move towards each other with velocity $\bm{V}_i(t)$ in a quiescent fluid of density $\rho$ and dynamic viscosity $\mu$. The fluid density and viscosity within the bubbles are denoted respectively $\zeta \rho$ and $\lambda \mu$ where $\zeta$ and $\lambda$ are the density and viscosity ratios, respectively.
The film, located between the two bubbles, is modeled using axisymmetric cylindrical polar coordinates $(r,z)$. The origin of the coordinate system $(O)$ is located at the center of the film. The film thickness is denoted by $h(r,t)$, while the shortest distance between the two bubbles is given by $d(t)$. From those definitions we have $V_2-V_1 = \dot{d} + \dot{a}_1 + \dot{a}_2$ where $V_1 = \bm{V}_1\cdot \bm{e}_z$ and $V_2 = \bm{V}_2\cdot \bm{e}_z$. As the velocities of the bubbles may vary with time, the frame of reference is not inertial in the most general configuration. The position of the two bubble surfaces is given by $z = h_1(r,t)$ and $z = h_2(r,t)$ (Figure \ref{fig:scheme}).

\subsection{Governing equations}

We assume that the flow within the film satisfies the Stokes equations. The Stokes flow solution provides a good approximation to the flow field near the bubbles when the Reynolds number, denoted as $Re=\rho \dot{d} \bar{a}/\mu$ is much less than unity \citep{kim1991}. Specifically, when $Re \ll 1$, the advection term in the Navier-Stokes equation, scaling as $\rho \dot{d}^2/\bar{a}$, is significantly smaller compared to the viscous term, which scales as $\mu \dot{d}/\bar{a}^2$. Here, $\bar{a}$ represents the reduced radius defined as $\bar{a}=a_1a_2/(a_1+a_2)$. The fluid motion is unsteady as the separating distance between the bubbles continuously changes. Nevertheless, if we consider the time scale to be on the order of $\bar{a}/\dot{d}$, the acceleration term becomes negligible compared to the viscous terms under the condition that $Re \ll 1$. Under this condition, we may disregard the acceleration of the film center relative to a stationary frame of reference.

The axisymmetric Stokes equations written in cylindrical coordinates read
\begin{align}
\frac{1}{r}\frac{\partial}{\partial r}(r u)+\frac{\partial w}{\partial z} &= 0 \label{eq:cdm}\\
0 &= -\frac{\partial p}{\partial r} + \mu \left(\frac{\partial ^2 u}{\partial r^2} + \frac{\partial ^2 u}{\partial z^2} + \frac{1}{r}\frac{\partial u}{\partial r}-\frac{u}{r^2}\right) \label{eq:qdmr}\\
0 &= -\frac{\partial p}{\partial z} + \mu \left(\frac{\partial  ^2 w}{\partial r^2} + \frac{\partial ^2 w}{\partial z^2} + \frac{1}{r}\frac{\partial w}{\partial r}\right)  \label{eq:qdmz}
\end{align}
where $u$ is the radial velocity, $w$ the axial velocity and $p$ the fluid pressure.
Equations \eqref{eq:cdm}-\eqref{eq:qdmz} are also valid within the bubbles (replacing $\rho$ by $\zeta\rho$ and $\mu$ by $\lambda \mu$ and adding a subscript $i$ to all the fields) provided that the condition $Re \ll \lambda$ is satisfied.

Before specifying the boundary conditions on the interface we define the tangential and normal unit vectors $\bm{n}_i$ and $\bm{t}_i$ as 

\begin{align}
\bm{n}_i = \left[1+\left(\frac{\partial h_i}{\partial r}\right)^2\right]^{-1/2}\left(\frac{\partial h_i}{\partial r} \bm{e}_r-\bm{e}_z\right), \quad
\bm{t}_i = \left[1+\left(\frac{\partial h_i}{\partial r}\right)^2\right]^{-1/2}\left(\bm{e}_r+\frac{\partial h_i}{\partial r}\bm{e}_z\right),
\label{eq:nt}
\end{align}
Note that $\bm{n}_2$ is pointing outward to the bubble surface while $\bm{n}_1$ is pointing inward. At the surface of bubble $i$ the impermeability condition reads

\begin{align}
\bm{u}\cdot\bm{n}_i = (-1)^i \dot{a}_i+\bm{V}_i\cdot\bm{n}_i \quad \text{on $z= h_i$}.
\end{align}

The tangential stress boundary condition may be expressed as $\bm{e}:(\bm{t}_i\otimes\bm{n}_i) = \bm{e}_i:(\bm{t}_i\otimes\bm{n}_i)$ on $z= h_i$ where $\bm{e}$ is the strain-rate tensor within the liquid and $\bm{e}_i$ the strain-rate tensor within bubble $i$. Except in the final section, the specific details regarding the internal flow within the bubble will not be taken into consideration.
Hence, for simplicity, we denote $\bm{e}_i:(\bm{t}_i\otimes\bm{n}_i) = f_i$, where $f_i(r,t)$ is the tangential stress exerted by the gas in the bubble $i$ on the interface.
Using \eqref{eq:nt} the tangential stress boundary condition reads
\begin{align}
&\mu\left[1+\left(\frac{\partial h_i}{\partial r}\right)^2\right]^{-1}\left[2\frac{\partial h_i }{\partial r } \left(\frac{\partial u}{\partial r}-\frac{\partial w}{\partial z} \right) + \left(\left(\frac{\partial h _i }{\partial r }\right) ^2 -1 \right)\left(\frac{\partial u}{\partial z}+\frac{\partial w}{\partial r} \right) \right] \nonumber \\
&= f_i \quad \text{on $z= h_i$}.
\label{eq:shear}
\end{align}

The assumption of spherical shapes is fulfilled as long as surface tension effects are much larger than dynamic normal stresses. This may be rationalized using the normal stress boundary conditions: $-p+2\mu\bm{e}:(\bm{n}_i\otimes\bm{n}_i) = -p_i+2\mu\bm{e}_i:(\bm{n}_i\otimes\bm{n}_i) + \gamma \kappa _i$ on $z= h_i$ where $\gamma$ is the surface tension and $\kappa _i$ the mean curvature of interface $i$. This boundary condition will be used to demonstrate \textit{a posteriori} the conditions under which the spherical assumption remains valid. As before, we define a function $g_i$ equals to the dynamic normal stress exerted by the bubble on the interface: $g_i (r,t) = -p_i+2\mu\bm{e}_i:(\bm{n}_i\otimes\bm{n}_i)$. Therefore,
\begin{align}
&-p + 2\mu \left[1+\left(\frac{\partial h_i}{\partial r}\right)^2\right]^{-1}\left(\left(\frac{\partial h _i }{\partial r }\right) ^2\frac{\partial u}{\partial r} - \frac{\partial h _i }{\partial r }\left(\frac{\partial u}{\partial z}+\frac{\partial w}{\partial r} \right)+\frac{\partial w}{\partial z}\right) \nonumber \\
&= g_i + \gamma \kappa _i \quad \text{on $z= h_i$} \label{eq:normal}.
\end{align}
The force on the $i$-th bubble reads

\begin{equation}
\bm{F}_i = (-1)^i\int _{S_i} (-p \bm{I} +2\mu \bm{e})\cdot \bm{n}_i dS
\end{equation}
where $S_i$ is the surface of bubble $i$ and $\bm{I}$ the identity matrix.
 Since the motion is axisymmetric, the only nonzero components of the force lies in the $z$ direction. From equation \eqref{eq:nt} we obtain for the force on the $i$-th bubble 
\begin{equation}
F_i =  2\pi (-1)^i \left[\int_0^{a_i}\left(p-2\mu \frac{\partial w}{\partial z}\right)rdr + \mu \int_0^{a_i}\left(\frac{\partial u}{\partial z}+ \frac{\partial w}{\partial r}\right) \frac{\partial h_i}{\partial r}rdr \right] \label{eq:force}
\end{equation}

\section{Lubrication theory}
\label{sec:lubrication}

In the subsequent section, we shall derive the lubrication equations. Although similarities exist between our derivation and prior investigations in the literature \citep{chesters1982,howell1996,savva2009}, a comprehensive derivation for the specific problem under consideration has not been previously done. Moreover, the previous derivations in the literature presuppose shear-free boundaries at the bubble surface, while in the present approach, the shear-free assumption will be justified \textit{a posteriori}. 
The assumption of shear-free interface is of primary importance since the set of lubrication equations between shear-free interfaces differs significantly from the standard lubrication equations \citep{davis1989,leal2007,michelin2019}. In the latter case, the flow is resisted either by the no-slip boundary condition at one of the interfaces \citep{leal2007,michelin2019} or by the flow inside one of the drops \citep{davis1989}. On the other hand, when both interfaces are free, the flow is resisted by the extensional effective viscosity \citep{howell1996}.

We consider the thin film limit, in which $d/\bar{a} = \epsilon ^2\ll 1$. Close to the axis of symmetry, i.e., for $r \ll \bar{a}$, we have $h_1 \sim - 1/2(d+r^2/a_1)$ and $h_2 \sim 1/2(d+r^2/a_2)$. Since $h=h_2-h_1$ we obtain
\begin{equation}
h(r,t) = d(t)+\frac{r^2}{2\bar{a}(t)} + \mathcal{O} \left(\frac{r^4}{a^4}\right)
\label{eq:h}
\end{equation}
Therefore in the film region, $r$ scales as $\sqrt{\bar{a}d}$ \citep{davis1989}. 
The appropriate normalizations for the lengths within the gap are $r=\sqrt{\bar{a}d} r^*$, $z=dz^*$ and $h = d h^*$, and the starred quantities are dimensionless. Equation \eqref{eq:h} written in non-dimensional form reads

\begin{equation}
h^*(r^*) = 1+\frac{r^{*2}}{2} + \mathcal{O} \left(\epsilon ^2\right)
\label{eq:h_adim}
\end{equation}

We non-dimensionalize the axial velocity as follows $w=\dot{d}w^*$. We get $u=u^*\dot{d}/\epsilon $ from the incompressibility condition. From the momentum equation in the $z$-direction, where viscous and pressure forces are in
balance, we obtain $p = p^*\mu \dot{d}/d$. Within the bubble the radial and axial length scales are $\sqrt{\bar{a}d}$ \citep{davis1989}. Hence, $f_i = f_i^*\lambda \mu \dot{d}/d $ and $g _i= g_i^*\lambda \mu \dot{d}/d $. Since the bubbles are spherical, we normalize the mean curvature by the inverse of $\bar{a}$ such that $\kappa = 1/\bar{a} \kappa^*$. We normalize the velocities as $\dot{a}_i = \dot{a}_i^*\dot{d}$ and $V_i = V_i^*\dot{d}$. 
 After this non-dimensionalization, the Navier-Stokes equations \eqref{eq:cdm} - \eqref{eq:qdmz} take the form


\begin{align}
\frac{1}{r^*}\frac{\partial}{\partial r^*}(r^* u^*)+\frac{\partial w^*}{\partial z^*} &= 0, \label{eq:cdm_adim}\\
0 &= -\epsilon^2\frac{\partial p^*}{\partial r^*} +\epsilon^2\frac{\partial ^2 u^*}{\partial r^{*2}} + \frac{\partial ^2 u^*}{\partial z^{*2}} + \epsilon^2\frac{1}{r^*}\frac{\partial u^*}{\partial r^*}-\epsilon^2\frac{u^*}{r^{*2}}, \label{eq:qdmr_adim}\\
0 &= -\frac{\partial p^*}{\partial z^*} +\epsilon^2\frac{\partial ^2 w^*}{\partial r^{*2}} + \frac{\partial ^2 w^*}{\partial z^{*2}} +\epsilon^2 \frac{1}{r^*}\frac{\partial w^*}{\partial r^*}, \label{eq:qdmz_adim}
\end{align}
The impermeability condition reads

\begin{align}
\epsilon^2\frac{\partial h_i^*}{\partial r^*}u^* - w^* = (-1)^i \dot{a}_i^* \left(1+\epsilon^2\left(\frac{\partial h_i^*}{\partial r^*}\right)^2\right)-\bm{V}_i^* \quad \text{on $z^*= h_i^*$}.
\end{align}
while the dynamic boundary conditions give
\begin{align}
&\left[1+\epsilon^2\left(\frac{\partial h_i^*}{\partial r^*}\right)^2\right]^{-1}\left[2\epsilon^2\frac{\partial h_i^* }{\partial r ^*}   \left(\frac{\partial u^*}{\partial r^*}-\frac{\partial w^*}{\partial z^*} \right) + \left(\epsilon^2\left(\frac{\partial h _i ^*}{\partial r ^*}\right) ^2 -1 \right)\left(\frac{\partial u^*}{\partial z^*}+\epsilon^2\frac{\partial w^*}{\partial r^*} \right)\right] \nonumber\\
&= \epsilon \lambda f_i^* \quad \text{on $z^*= h_i^*$} \label{eq:shear_adim},\\
&-p^* + 2\left[1+\epsilon^2\left(\frac{\partial h_i^*}{\partial r^*}\right)^2\right]^{-1}\left(\epsilon^2\left(\frac{\partial h _i^* }{\partial r ^*}\right) ^2\frac{\partial u^*}{\partial r^*} - \frac{\partial h _i ^*}{\partial r^* }\left(\frac{\partial u^*}{\partial z^*}+\epsilon^2\frac{\partial w^*}{\partial r^*} \right)+\frac{\partial w^*}{\partial z^*}\right)  \nonumber \\
&= \lambda g_i^* + \frac{\epsilon ^2}{Ca} \kappa _i^* \quad \text{on $z= h_i^*$},
\label{eq:normal_adim}
\end{align}
where $Ca = \mu \dot{d}/\gamma$ is the capillary number. As it can be seen from equation \eqref{eq:normal_adim}, the deformation relative to the initial spherical shapes remains small provided that $Ca \ll \epsilon ^2$. 


We seek solutions of this set of equations in the form of expansions in powers of the small parameter $\epsilon ^2$ \citep{howell1996}: $u^* = u_0^* + \epsilon^2 u_2^* + ...$, $w^* = w_0^* + \epsilon^2 w_2^* + ...$,  $p^* = p_0^* + \epsilon^2 p_2^* + ...$. 
Inserting this expansion in Equations \eqref{eq:cdm_adim} - \eqref{eq:qdmz_adim}, we obtain to the lowest order

\begin{align}
\frac{1}{r^*}\frac{\partial}{\partial r^*}(r^* u_0^*)+\frac{\partial w_0^*}{\partial z^*} &= 0, \label{eq:cdm_adim0}\\
0 &= \frac{\partial ^2 u_0^*}{\partial z^{*2}}, \label{eq:qdmr_adim0}\\
0 &= -\frac{\partial p_0^*}{\partial z^*} + \frac{\partial ^2 w_0^*}{\partial z^{*2}}. \label{eq:qdmz_adim0}
\end{align}
The impermeability condition at the bubble surface is given by
\begin{align}
- w_0^* = (-1)^i \dot{a}_i^* -\bm{V}_i^* \quad \text{on $z^*= h_i^*$}. \label{eq:kinematic0}
\end{align}
Assuming that the viscosity ratio is small, \textit{i.e.} $\lambda \sim \epsilon$ the tangential boundary condition \eqref{eq:shear_adim} gives 

\begin{equation}
\frac{\partial u _0^*}{\partial z^*} = 0 \quad \text{on $z^*= h_i^*$} \label{eq:shear_adim0}.
\end{equation}
Making use of Equations \eqref{eq:qdmr_adim0} and \eqref{eq:shear_adim0}, we obtain $u _0^*\equiv u_0^*(r^*)$. Hence, the radial velocity is uniform along the film. As a consequence, equation \eqref{eq:cdm_adim0} and $w_0^*(r^*,z^*=0) =0$ yields
\begin{equation}
 w_0^*(r^*,z^*) = - \frac{1}{r^*}\frac{\partial}{\partial r^*}(r^* u_0^*)z^*.
\label{eq:w0}
\end{equation}
Substitution into \eqref{eq:qdmz_adim0} yields $p _0^* \equiv  p_0^*(r^*)$. We now insert the $\mathcal{O}(\epsilon ^2)$ terms in the expansions in equation \eqref{eq:qdmr_adim} to obtain an equation for the radial velocity 
\begin{equation}
0 = -\frac{\partial p_0^*}{\partial r^*} + \frac{\partial^2 u_0^*}{\partial r^{*2}} + \frac{\partial^2 u_2^*}{\partial z^{*2}} + \frac{1}{r^*}\frac{\partial u_0^*}{\partial r^*}-\frac{u_0^*}{r^{*2}}. \label{eq:qdmr2}
\end{equation}
Similarly, by inserting the $\mathcal{O}(\epsilon ^2)$ terms in the tangential stress boundary conditions \eqref{eq:shear_adim} we obtain
\begin{equation}
2\frac{\partial h_i^* }{\partial r^* }  \left(\frac{\partial u_0^*}{\partial r^*}-\frac{\partial w_0^*}{\partial z^*} \right) -\left(\frac{\partial u_2^*}{\partial z^*}+\frac{\partial w_0^*}{\partial r^*} \right) = \frac{\lambda}{\epsilon}f_i^* \quad \text{on $z^*= h_i^*$} \label{eq:shear_adim2}.
\end{equation}
Upon integrating equation \eqref{eq:qdmr2} from $h_1^*$ to $h_2^*$ and making use of equations \eqref{eq:w0} and \eqref{eq:shear_adim2} we find

\begin{equation}
0 = - \frac{\partial p_0^*}{\partial r^*} + 2\left[\frac{\partial}{\partial r^*}\left(\frac{1}{r^*}\frac{\partial}{\partial r^*}(r^* u_0^*)\right) + \frac{1}{h^*}\frac{\partial h^*}{\partial r ^*}  \left(2\frac{\partial u_0^*}{\partial r^*}+\frac{u_0^*}{r^*} \right)\right] - \frac{\lambda}{\epsilon h^*}(f_2^*(r,t)  - f_1^*(r,t)).
\label{eq:qdm_shear}
\end{equation}

Then, under the assumption of negligible shear stress at the interfaces ($\lambda \ll \epsilon$), the last equation may be written 
\begin{equation}
0  = - \frac{\partial p _0^*}{\partial r^*} + 2 \left[\frac{\partial}{\partial r^*}\left(\frac{1}{r^*}\frac{\partial}{\partial r^*}(r^* u_0^*)\right) + \frac{1}{h^*}\frac{\partial h^*}{\partial r^* }  \left(2\frac{\partial u_0^*}{\partial r^*}+\frac{u_0^*}{r^*} \right)\right] \label{eq:qdm_final},
\end{equation}
Equations \eqref{eq:h_adim}, \eqref{eq:cdm_adim0}, \eqref{eq:kinematic0} and \eqref{eq:qdm_final} constitute a system of lubrication equations that describe
the flow within the thin gap located between the two spherical bubbles. 
The main limitations of those equations are their limited range of validity: $Re \ll 1$, $Ca \ll \epsilon ^2$, and $\lambda \ll \epsilon$. We will drop the subscript 0 for convenience in the next sections.

\section{Forces on the bubble}
\label{sec:force}

Each of the two integrals on the right-hand side of \eqref{eq:force} may be decomposed over an inner region where lubrication assumption is valid and an outer region on the scale of the bubble \citep{cox1967}. More precisely, we may define $R_\infty$ such that $r = R_\infty$ may be considered lying between the inner and outer regions. Hence, $R_\infty$ should be in the range $\sqrt{\bar{a}d} \ll R_\infty \ll \bar{a}$ \citep{davis1989}. Because of the somewhat arbitrary manner in which $R_\infty$ is defined, the final expression of the force should not depend on this parameter. It will naturally disappear from the total force when adding the outer solution to the inner lubrication region \citep{cox1967,oneill1967}. 
In the present configuration, the outer solution is known analytically for $Re=0$ \citep{haber1973}. Nevertheless, a detailed knowledge of the outer solution is not required by restricting our attention to the singular terms of the expansion, which may be obtained from the inner solution alone. 


Using the normalizations defined in section \ref{sec:lubrication} and expressing the force on the $i$-th bubble as $F_i = \mu \dot{d} \bar{a}F_i^*$ and $R_\infty= \bar{a} R_\infty^*$ the inner force contribution reads  


\begin{equation}
F_i^{i*} = 2\pi (-1)^i \left[\int_0^{R_\infty^*/\epsilon}\left(p^*-2\frac{\partial w^*}{\partial z^*}\right)r^*dr^* + \int_0^{R_\infty^*/\epsilon}\left(\frac{\partial u^*}{\partial z^*}+ \epsilon ^2\frac{\partial w^*}{\partial r^*}\right) \frac{\partial h_i^*}{\partial r^*}r^*dr^*. \right] \label{eq:forcen}
\end{equation} 
Since $F_1^{i*} = -F_2^{i*}$, we denote $F^*$ the inner force on the first bubble for convenience. 
By substituting the expansions for $u^*$, $w^*$, and $p^*$ in terms of the small parameter in equation \eqref{eq:forcen} and recognizing that, at the lowest order, $u^*$ is independent of $z^*$, we may write 

\begin{equation}
F^* = -2\pi \int_0^{R_\infty^*/\epsilon}\left(p^*-2\frac{\partial w^*}{\partial z^*}\right)r^*dr^* \label{eq:forcen2}
\end{equation} 
Hence, to determine the force, it is necessary to compute both the velocity profile and the pressure distribution within the film.
One may also note that in contrast to \citet{davis1989} and \citet{leal2007}, the pressure and viscous stress in our analysis have comparable magnitudes in Equation \eqref{eq:forcen2} due to the assumption of shear-free interfaces \citep{savva2009}.

\subsection{Viscous stress and pressure distribution}
\label{sec:stress}

To obtain the radial velocity profile we integrate \eqref{eq:cdm_adim0} with
respect to $z^*$

\begin{equation}
\int_{h_1^*}^{h_2^*}\frac{1}{r^*}\frac{\partial}{\partial r^*}(r^* u^*)dz^*+w^*(h_2^*)-w^*(h_1^*) = 0.
\end{equation}
If we apply the boundary conditions for $w^*$ from \eqref{eq:kinematic0} we obtain the radial velocity

\begin{equation}
u^*(r^*)= -\frac{1}{2}\frac{r^*}{h^*(r^*)},
\label{eq:u}
\end{equation}
where $h^*$ is given by \eqref{eq:h_adim}.
Substitution into \eqref{eq:cdm_adim0} thus yields 
\begin{equation}
\frac{\partial w^*}{\partial z^*}(r^*) =  \frac{1}{h^{*2}(r^*)}.
\label{eq:viscous_stress}
\end{equation}

\begin{figure}
\centering
\includegraphics[width=5.5cm]{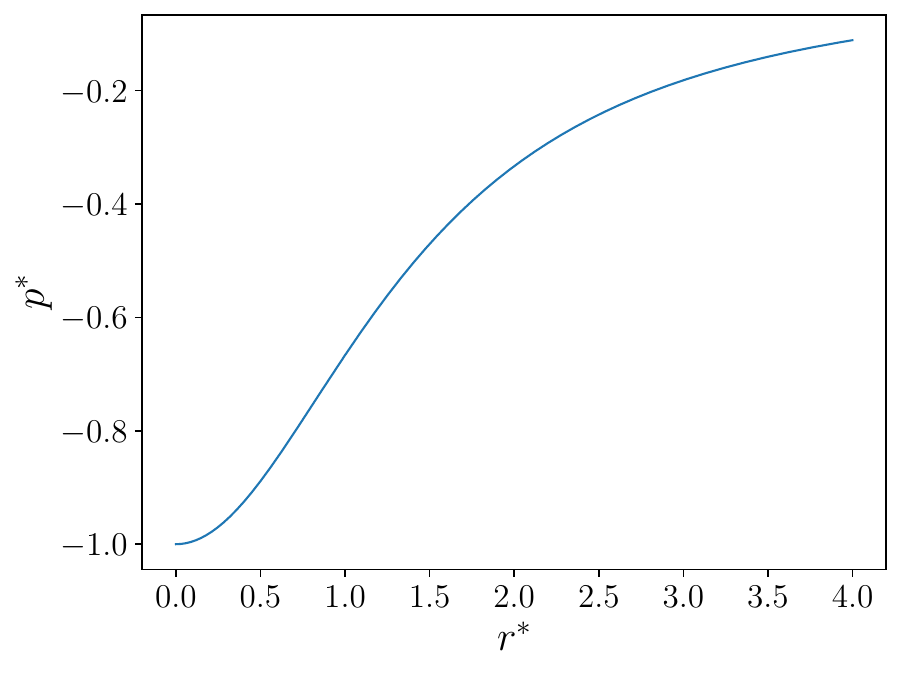}\includegraphics[width=5.5cm]{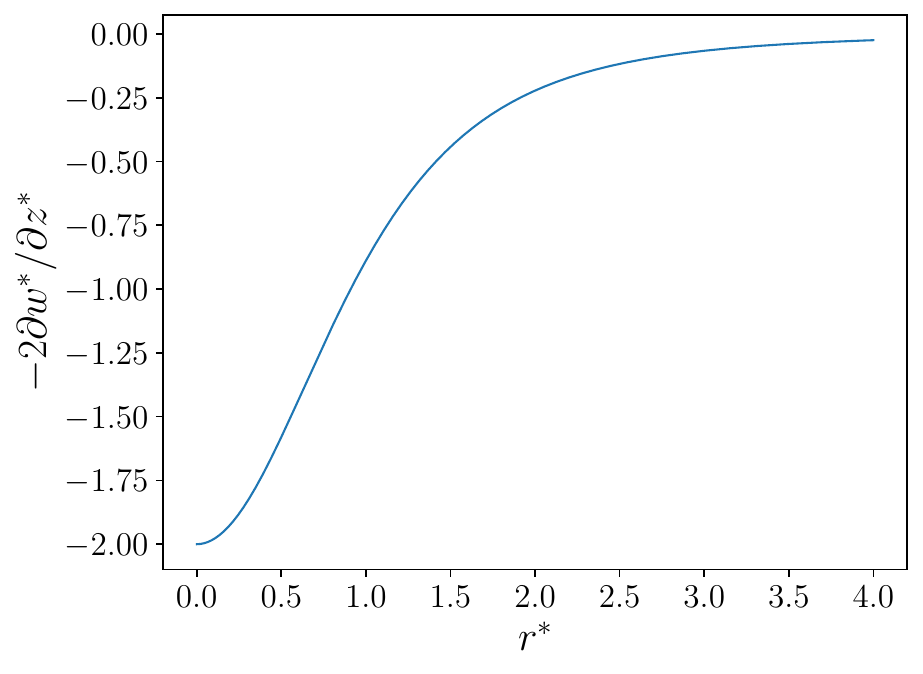}\\
(a) \hspace{5cm} (b)
\caption{The dimensionless pressure and normal viscous stress. (a): Pressure. (b) : Normal viscous stress.}
\label{fig:pressure}
\end{figure}

Integration of Equation \eqref{eq:qdm_final} in $r^*$ (see appendix \ref{app:pressure} for the details), and making use of Equation \eqref{eq:u} the pressure profile reads 

\begin{equation}
p^*(r^*)  = - \frac{1}{h^*(r^*)} + p_\infty ^*
\label{eq:pressure}
\end{equation}
where $p_\infty^*$ is a $\mathcal{O}(1)$ term related to the pressure outside the film. Without loss of generality, we assume that $p_\infty ^* = 0$. Both the pressure \eqref{eq:pressure} and the normal viscous stress  \eqref{eq:viscous_stress} combine to form the overall stress within the thin film. These quantities are singular in the limit of small thickness as they scale proportionally to $1/h^*$ or $1/h^{*2}$, respectively. Figure \ref{fig:pressure} illustrates their behavior. Both quantities approach zero as the radial distance $r$ approaches infinity. 

\subsection{Lubrication forces}
The innner contribution to the force on the bubble can be obtained by inserting Equations \eqref{eq:viscous_stress} and \eqref{eq:pressure} in \eqref{eq:forcen2}. The calculation details are summarized in appendix \ref{app:force}. This yields,


\begin{equation}
F^* = -4\pi\log\epsilon + \mathcal{O}(1).
\label{eq:forcea}
\end{equation}
Note that the viscous stress contributes only to the $\mathcal{O}(1)$ term. 
The preceding formula may be rewritten in dimensional form as
\begin{equation}
F = -4\pi\mu \bar{a} \dot{d} \log \epsilon
\label{eq:forced}
\end{equation}
where we recall that $\dot{d} = V_2-V_1 - (\dot{a}_1 +\dot{a}_2)$.
The formula \eqref{eq:forced} has the advantage of being applicable for drops with different radii as long as $Re \ll 1$, $Ca \ll \epsilon ^2$ and $\lambda \ll \epsilon$. However, since it is a weak logarithm singularity, its accuracy is low except for very thin film, potentially violating the conditions $Ca \ll \epsilon ^2$ and $\lambda \ll \epsilon$ in practical conditions. Hence, there is a need to derive the $\mathcal{O}(1)$ terms to enhance solution accuracy. These terms may be obtained by properly matching the inner solution with the outer solution \citep{oneill1967}. Instead, we use previous findings from the literature to compute them. Additionally, given that the $\mathcal{O}(1)$ term depends on the specific configuration under consideration, we focus on cases involving bubbles with identical radii either moving at identical velocities or experiencing a constant rate of change in radius. We also consider the case of a bubble translating or expanding/draining near a plane-free surface.

\subsection{Determination of the $\mathcal{O}(1)$ contribution to the force}
\label{sec:viscf}

\begin{figure}
\centering
\includegraphics[width=6cm]{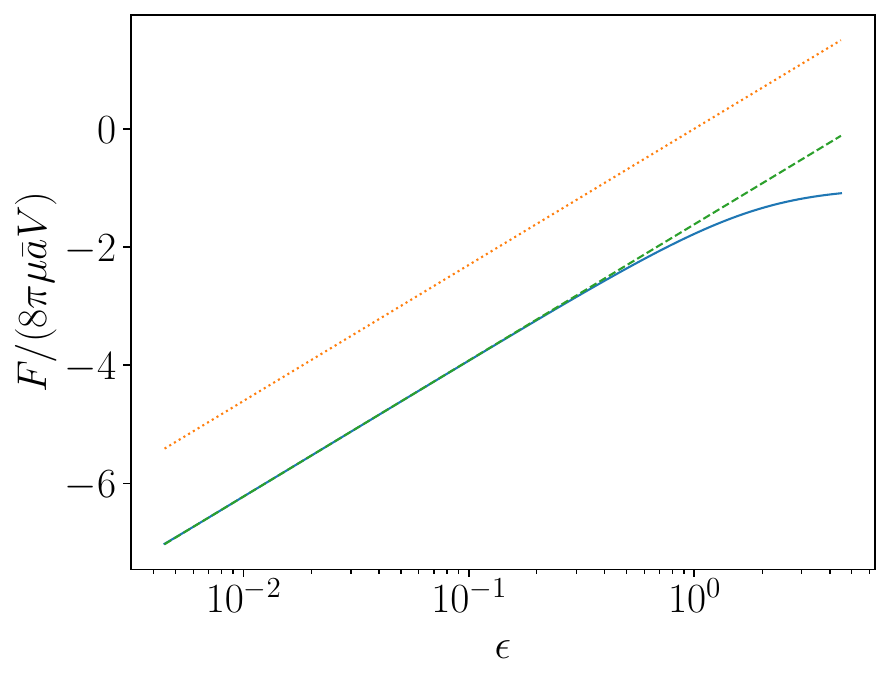}
\caption{Dimensionless force resisting the motion of two translating identical bubbles. \color{blue_py}--- \color{black} : Exact bi-spherical coordinate solution (Equation \eqref{eq:bispherical_trans} for $\lambda=0$),\color{orange_py}$\cdot \cdot$ \color{black}: lubrication solution $F = 8\pi\mu \bar{a} V \log \epsilon$, \color{green_py}- - \color{black} : equation \eqref{eq:kim_A}.}
\label{fig:kim}
\end{figure}

We first consider the case of two identical bubbles translating with opposite and equal velocity $V$ such that $\dot{d} = -2V$. 
Formula \eqref{eq:forced} may be expressed as $F = 8\pi\mu \bar{a} V \log \epsilon$. \citet{kim1991} derived an asymptotic formula based on the bi-spherical coordinate solution proposed by \citet{haber1973}. Their solution may be expressed as 

\begin{equation}
F = 8\pi\mu \bar{a} V \left(\log \epsilon + A\right)
\label{eq:kim_A}
\end{equation}
where $A = -\gamma -3/2\log 2$ and $\gamma$ is the Euler's constant. We recall that $\bar{a}$ denotes the reduced radius, which is half the radius of the bubbles in this case, and $\epsilon$ is defined as $\epsilon = (d/\bar{a})^{1/2}$. Figure \ref{fig:kim} displays the variation of force with respect to film thickness. 
The formula \eqref{eq:kim_A} (depicted with dashed line) is in good agreement with the bi-spherical coordinate solution (depicted with continuous line) up to $\epsilon \approx 1$. In contrast, the solution derived solely from the inner lubrication region (dotted line) proves inaccurate even for small $\epsilon$.

We now consider the case of a bubble translating toward a free surface with a velocity $V$. 
In this case, Equation \eqref{eq:forced} reads $F = 4\pi\mu \bar{a} V \log \epsilon$. Additionally, solution \eqref{eq:kim_A} may be readily extended to the case of a free surface by noting that the plane of symmetry separating the two bubbles is a shear-free boundary. By substituting $\bar{a}$ with $\bar{a}/2$ and $d$ with $2d$ in \eqref{eq:kim_A}, we obtain

\begin{equation}
F = 4\pi\mu \bar{a} V \left(\log \epsilon + A_\infty\right)
\label{eq:kim_planar}
\end{equation}
where $A_\infty = -\gamma -\log (2)/2$.

\begin{figure}
\centering
\includegraphics[width=6cm]{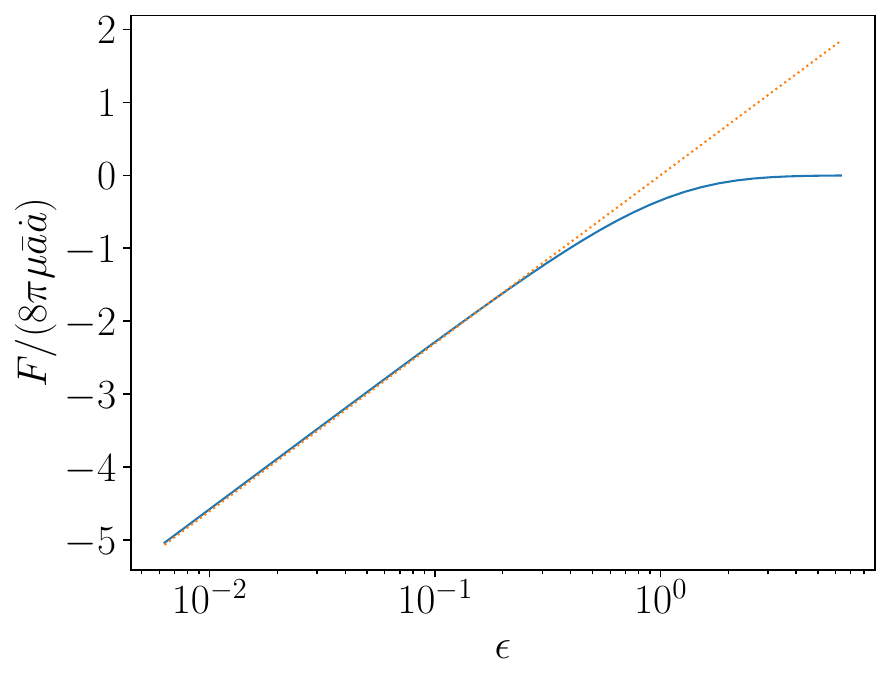}
\caption{Dimensionless force resisting the motion of two growing bubbles in viscous flow. \color{blue_py}--- \color{black} : Exact bi-spherical coordinate solution (Equation \eqref{eq:bispherical}), \color{orange_py}$\cdot \cdot$ \color{black}:  lubrication solution $8\pi\mu \bar{a} \dot{a} \log \epsilon$. } 
\label{fig:force_visc}
\end{figure}

To the best of our knowledge, the contribution arising from the variation of the bubble radius in formula \eqref{eq:forcea} has not been derived so far. To verify the accuracy of this result, the bi-spherical coordinate solution proposed in \citet{michelin2018} can be used. However, it is important to note that \citet{michelin2018} did not explicitly express the force acting on the bubble, instead requiring the solution of a linear system comprising four equations. In the specific scenario of two identical bubbles, the system's complexity can be reduced, allowing for the derivation of an explicit form for the force, as demonstrated in Appendix \ref{app:visc}. 
Denoting $\dot{a}$ the time derivative of the bubble radius, we have $\dot{d} = -2\dot{a}$. Formula \eqref{eq:forced} can be expressed as $F = 8\pi\mu \bar{a} \dot{a} \log \epsilon$. Consequently, similar to the case of purely translating bubbles, one can seek the force expression as

\begin{equation}
F = 8\pi\mu \bar{a} \dot{a} \left(\log \epsilon + B\right)
\label{eq:michelin}
\end{equation}
where $B$ represents a term of order $\mathcal{O}(1)$, which may be fitted from the bi-spherical coordinate solution. In our specific scenario, the best-fit yields $B \approx 0$. Indeed, as depicted in Figure \ref{fig:force_visc}, there is an excellent agreement between the bi-spherical coordinate solution (depicted with a continuous line) and the predictions of lubrication theory (depicted with a dotted line).
Formula \eqref{eq:michelin} may be readily extended to the case of a bubble with a time-dependent radii near a free surface. In this case, the force may be expressed as

\begin{equation}
F = 4\pi\mu \bar{a} \dot{a} \left(\log \epsilon + B_\infty\right)
\label{eq:michelin2}
\end{equation}
where $B_\infty \approx \log 2 $.

\section{Some applications to bubble dynamics}
\label{sec:examples}
In this section, we explore how our findings can be used to determine the dynamics of translating bubbles with time-dependent radii. We first consider the case of a bubble translating toward a free surface under the action of a constant body force. Subsequently, we analyze the dynamics of two expanding bubbles in the absence of external forces.



\subsection{Bubble translating toward a free surface under the action of gravity}
\label{sec:vaka}

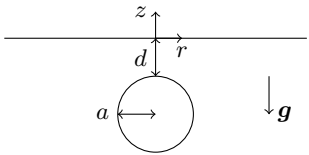
\begin{figure}
\centering
\begin{tikzpicture}[scale=1]


\draw [] (0,-1) circle (0.5);
\draw[<->] (0,-0.5) -- node[left] {$d$}(0,0) ;

\draw[] (-2,0) -- (2,0);
\draw[->] (0,0) --(0,0.35) node [left] {$z$};
\draw[->] (0,0) --(0.35,0) node [below] {$r$};

\draw[->] (1.5,-0.5) --(1.5,-1) node [right] {$\bm{g}$};
\draw[<->] (0,-1) -- (-0.5,-1) node[left] {$a$};

\end{tikzpicture}
\caption{Spherical bubble approaching a flat free surface.}
\label{fig:scheme_free}
\end{figure}

We consider a bubble of radius $a$ translating toward a free surface under the action of gravity denoted by $g$ (Figure \ref{fig:scheme_free}). Our primary focus is determining the film drainage time, which is the duration required for the bubble to drain the film between its upper surface and the free surface.
A significant limitation of the current analysis, which restricts its practical application, is the assumption of negligible deformations of the free surface and of the bubble shape relative to a spherical geometry, which is expressed as $Ca \ll \epsilon ^2$. 
This assumption necessitates extremely small bubbles and high surface tension, as demonstrated later in this section.

We are unaware of any experimental results measuring the film drainage time scale for small bubbles in viscous flow except the recent experiments of \citet{vakarelski2018}. The absence of prior research can be attributed to two reasons. First, it is very difficult experimentally to avoid the contamination of the interface with impurities \citep{vakarelski2018}. Second, tracking of bubbles smaller than approximately 200 $\mu m$ requires high-speed video cameras equipped with microscope \citep{vakarelski2018}. \citet{vakarelski2018} studied the free rise and coalescence of small air-bubbles ($a \geq 100 \mu m$) at a liquid-air interface. They use a fluorocarbon liquid with a density of $\rho = 2030$ kg.m$^{-3}$ and a viscosity of $\mu = 0.0192$ Pa.s, which is highly resistant to surface-active contamination. Their findings reveal that the coalescence time $t_c$ for bubbles with radii $125 \mu$m is approximately $3.6$ ms and increases for larger bubbles. 
In the following, we will demonstrate that our model agrees with their experimental results. Balancing the buoyancy force (neglecting the gas density and noting that $\bar{a}=a$) with equation \eqref{eq:kim_planar}, one obtain 

\begin{equation}
- \frac{d\epsilon^2}{dt^*}(\log \epsilon ^2 + 2 A_\infty) + \frac{2}{3} =0
\label{eq:force_balance}
\end{equation}
where $t = \mu /(\rho a g) t^*$. Integration of this equation with respect to time yields 

\begin{equation}
(1-2A_\infty )\epsilon^2 - \epsilon^2\log(\epsilon^2) = (1-2A_\infty )\epsilon^2(0) - \epsilon^2(0)\log(\epsilon^2(0))-\frac{2}{3}t^*
\end{equation}
Since the coalescence time $t_c$ is defined as the duration required for the film to reach a zero thickness we get

\begin{equation}
t_c^* = \frac{3}{2}\epsilon^2(0)\left( 1 -2A_\infty - \log(\epsilon^2(0))\right).
\label{eq:tc}
\end{equation}

One may note that in contrast to solid particles, the film drainage process occurs in a finite time. \citet{vakarelski2018} did not explicitly state the initial thickness $d(0)$ at which coalescence time measurements were conducted. However, they indicated that the time was determined once the bubble reached the interface. Given their spatial resolution exceeding $1.7\mu$m, it is reasonable to assume that $d(0)$ falls within the range of 2 pixels (3.4$\mu$m) to six pixels ($10.2\mu$m). 
\begin{figure}
\centering
\includegraphics[width=6cm]{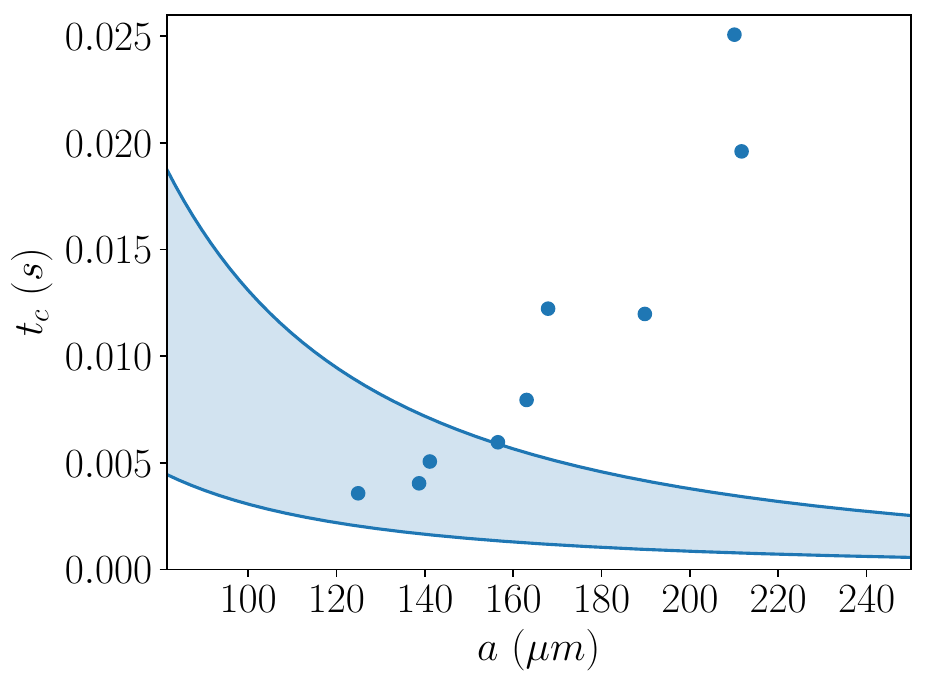}
\caption{Film drainage time versus bubble radius. \color{blue_py}$\bullet$\color{black}: \citet{vakarelski2018} experiments, coloured area: equation \eqref{eq:tc} for $3.4\mu$m $\leq d(0) \leq 10.2\mu$m.}
\label{fig:vakarelski}
\end{figure}
Equation \eqref{eq:tc} is in good agreement with the experimental findings for $a \leq 150 \mu$m (Figure \ref{fig:vakarelski}). However, it should be noted that our coalescence time prediction exhibits a behavior of $a^{-1}$ for a fixed initial film thickness $d(0)$, while \citet{vakarelski2018} observed an increase in coalescence time proportional to $a^2$ for larger bubbles (Figure \ref{fig:vakarelski}). This probably indicates that the results reported by \citet{vakarelski2018} for the smallest bubbles lie at the boundary between the Taylor regime of film drainage, where deformation is negligible, and the Reynolds regime, where deformation is non-negligible \citep{ivanov1999}.

Our analysis assumes that the conditions $Ca \ll \epsilon ^2$, $Re \ll 1$, and $\lambda \ll \epsilon$ are met. In the subsequent discussion, we address the validity of each of these assumptions. 
First, we have neglected (convective) inertia effects, which is acceptable since the Reynolds number associated with bubble motion is significantly less than unity. Indeed, to leading order, the velocity of the bubble reads $1/(3\log\epsilon)$. Hence, the assumption $Re \ll 1$ requires

\begin{equation}
Ar \ll \frac{\log \epsilon}{\epsilon}
\label{eq:Ar}
\end{equation}
where $Ar = \rho ^2 a^3 g/\mu ^2 $ is the Archimedes number. Typically, for $\epsilon = 0.1$, we find $|\log(\epsilon)|/\epsilon \approx 23$. Consequently, condition \eqref{eq:Ar} is met in the experiments conducted by \citet{vakarelski2018} for $a = 150 \mu m$, since $Ar \approx 0.4$. 
However, the effect of fluid unsteadiness on the force balance requires further consideration. From Equation \eqref{eq:force_balance}, we note that the timescale $\tau$ governing the balance between drag and buoyancy forces is $\mu/(\rho a g)\epsilon^2 \log(\epsilon)$. This result can also be derived from Equation \eqref{eq:tc}. The unsteady term in the Navier-Stokes equation becomes negligible compared to the viscous term only if $\tau \gg \rho a^2/\mu$ \citep{kim1991}, which requires

\begin{equation}
Ar \ll \epsilon^2 \log \epsilon
\label{eq:Ar_unsteady}
\end{equation}

This criterion is much more restrictive than \eqref{eq:Ar}. To relax this condition, one may consider the influence of unsteady forces in the viscous regime, such as added mass and history force, in the force balance (equation \eqref{eq:force_balance}). However, while the added mass as a function of the free surface distance may be obtained \citep{miloh1977}, to our knowledge, no expression exists for the history force of a bubble approaching a free surface. Furthermore, as illustrated in Figure \ref{fig:vakarelski} and as discussed in the original study by \citet{vakarelski2018}, the force balance \eqref{eq:force_balance} proves to be sufficiently accurate to predict the dynamics of small bubbles with radii up to approximately $100\mu$m. However, it must be said that in this case, the criterion \eqref{eq:Ar_unsteady} is not satisfied. This requires further investigation of the impact of unsteady forces on the dynamics of bubbles in proximity to a boundary.





We now consider the criterion $Ca \ll \epsilon ^2$. To leading order, the velocity of the bubble reads $1/(3\log\epsilon)$. 
Substituting this estimate into the Capillary number yields

\begin{equation}
Bo \ll \epsilon \log \epsilon
\end{equation}
where $Bo = \rho g a^2/\gamma$ is the Bond number. 
In the experiments of \citet{vakarelski2018}, when considering a bubble with a radius of $100\mu$m immersed in a liquid, the resulting Bond number is approximately $Bo \approx 9 \times 10^{-3}$, whereas for a bubble with a radius of $200\mu$m, $Bo \approx 3.6 \times 10^{-2}$. Subsequently, assuming significant deformation occurs when the Bond number approaches $\epsilon \log \epsilon$, for the $100\mu$m bubble, one might expect deformation when $d \approx a Bo ^2 \approx 8 \times 10^{-9}$m, and for the $200\mu$m bubble, $d \approx a Bo^2 \approx 2.6 \times 10^{-7}$m. Assuming that the critical thickness at which the film ruptures is typically around $10^{-8}$m \citep{chatzigiannakis2020}, we anticipate that the current analysis is applicable to bubbles with a radius of approximately $100\mu$m. Indeed, in such cases, deformation tends to occur at smaller film thicknesses. For larger bubbles the agreement between the model and the experiments is questionable since significant interface deformations are expected, indicating the possibility of an alternative regime of film drainage. Measurements of the film thickness of gas bubbles ascending towards a free surface under gravity were performed by \citet{kovcarkova2013}. They observed an exponential reduction in film thickness over time, with a characteristic time scale proportional to $\mu a / \gamma$ for small bubbles. For the present system with $a\approx 150 \mu$m, $\mu a / \gamma$ is approximately $0.1$ ms, indicating that the time scale for drainage after deformation is considerably shorter than the previously calculated value, providing confidence in the estimated coalescence time for $a \leq 150 \mu$m.

Finally, the condition $\lambda \ll \epsilon$ is fulfilled since the viscosity ratio is very small $\lambda\approx 10^{-3}$ in the experiments of \citet{vakarelski2018}. Indeed, the influence of shear stress within the bubble will manifest at film thicknesses smaller than the critical thickness at which film rupture occurs. However, one may expect a significant effect of the gas viscosity in the case of an air-water system ($\lambda \approx 0.018$). This topic will be discussed in the last section. 

\subsection{Dynamics of two growing bubbles}
\label{sec:ohashi}

Equation \eqref{eq:forcea} may be used in principle to predict the dynamics of two bubbles with time-dependent radii in the Stokes regime without external forces. A related analysis was conducted by \citet{michelin2019} for a bubble near a wall. Their findings indicated that the boundary conditions on the bubble surface determined whether the bubble would either continuously drain the fluid separating it from the wall or rebound once before draining the film. In their configuration, the translational lubrication force has a singularity that scales as $1/\epsilon^2$, which is significantly more singular than the inflating lubrication force (with $d$ fixed) that produced a weak logarithmic singularity. In contrast, in the present configuration, the translational lubrication force itself has a weak logarithmic singularity, while the inflating contribution to the lubrication force (with $d$ fixed) generates terms smaller than the order one, as the velocity $w$ involved in this case is proportional to $\epsilon ^2 \dot{a}$ is negligible. Consequently, when considering two force-free bubbles in the Stokes regime, Equation \eqref{eq:forcea} equals zero, leading to $\dot{d}=0$. This implies that the film thickness does not vary over time, 
which requires that $V_2 - V_1 = \dot{a}_1 + \dot{a}_2$. When bubbles approach each other ($V_2 - V_1 < 0$), the force vanishes if the bubbles shrink at half their relative velocity, for instance. Conversely, when bubbles move apart, the force vanishes if they expand at half their relative velocity. 

\begin{figure}
\centering
\includegraphics[width=6cm]{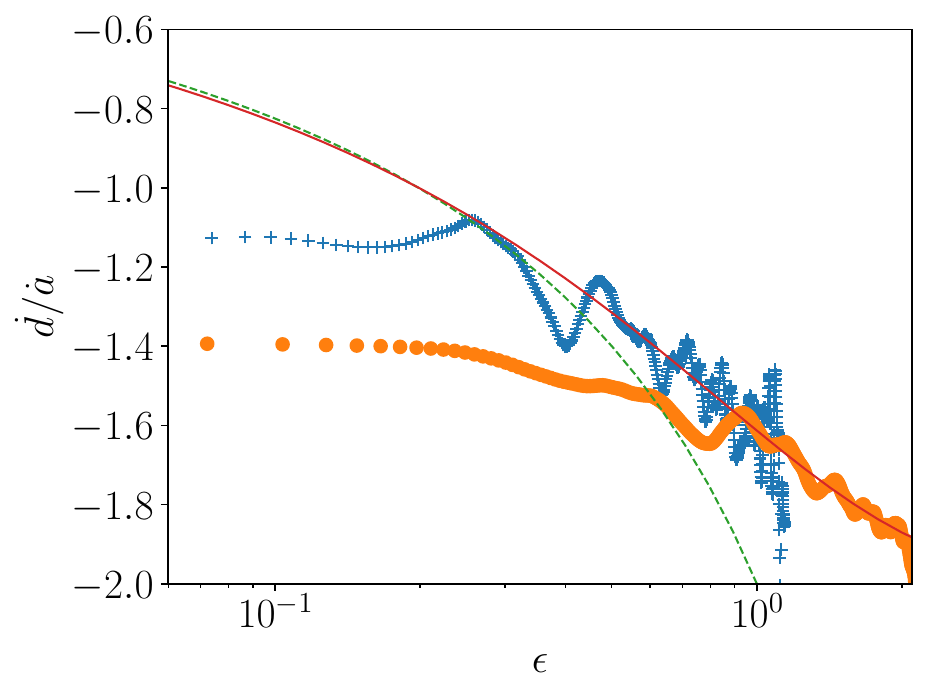}
\caption{Evolution of $\dot{d}/\dot{a}$ as a function of the film thickness. \color{blue_py}+\color{black} : Experimental results of \citet{ohashi2022} for $Ca_{\dot{a}} \approx 0.005$, \color{orange_py} $\bullet$ \color{black} : Experimental results of \citet{ohashi2022} for $Ca_{\dot{a}} \approx 0.1$, \color{green_py} - - \color{black} Equation \eqref{eq:h_0dot}, \color{red_py}--- \color{black} : bi-spherical coordinate solution based on equations \eqref{eq:bispherical_trans} and \eqref{eq:bispherical}. Since the radius, as well as the growth rate, are not exactly the same between the two bubbles in the experiments of \citet{ohashi2022} we compute the average value of the bubble radius.} 
\label{fig:ohashi}
\end{figure}

This result contrasts with those of a recent experimental investigation by \citet{ohashi2022}, who observed the decrease in time of the film thickness $d$ of two adjacent, expanding bubbles with low capillary numbers. To accurately describe the dynamics of two bubbles with time-dependent radii, it is necessary to include the $\mathcal{O}(1)$ contributions to both the inflating (or draining) and translating force. In the case of force-free bubbles with equal velocity $V$ and expanding rate $\dot{a}$ and to the linearity of the Stokes equation, we may sum equations and \eqref{eq:kim_A} and \eqref{eq:michelin}. This yields 
\begin{equation}
-(\log \epsilon + A)V + (\log \epsilon + B)\dot{a} = 0
\end{equation}
where $A$ and $B$ are defined in section \ref{sec:viscf}. Noting that $\dot{d}=2V - 2\dot{a}$ the force-free condition finally reads

\begin{equation}
\frac{\dot{d}}{\dot{a}} = \frac{2(B - A)}{\log(\epsilon) + A}
\label{eq:h_0dot}
\end{equation}
Equation \eqref{eq:h_0dot} indicates that the film thickness is a decreasing function of time for two growing bubbles in agreement with the experimental results of \citet{ohashi2022}. They consider the coalescence of two growing bubbles in highly viscous liquids ($5.10 ^{-6}\lesssim Re \lesssim 10 ^{-4}$, $2.10^{-5}\lesssim \lambda \lesssim 2.10^{-4}$). Two approximatively equal-sized bubbles were injected thanks to a syringe inside a closed box. Then, the pressure inside the box was reduced. 
We can go further in the comparison with \citet{ohashi2022} experiments by directly comparing $\dot{d}/\dot{a}$ to their experimental values. Specifically, we focus on experiments characterized by the smallest capillary numbers based on the growth rate, denoted as $Ca_{\dot{a}} = \mu \dot{a}/\gamma \approx 0.1$ and $Ca_{\dot{a}} \approx 0.005$, originally labeled as cases (a) and (c) in \citet{ohashi2022}. As depicted in Figure \ref{fig:ohashi}, there is excellent agreement between the force-free bi-spherical solution (obtained by equating equations \eqref{eq:bispherical_trans} and \eqref{eq:bispherical}) and the experimental findings of \citet{ohashi2022} up to $\epsilon \approx 1$ for $Ca_{\dot{a}} \approx 0.1$ and up to $\epsilon \approx 0.2$ for $Ca_{\dot{a}} \approx 0.005$. However, for thinner films, noticeable bubble deformation occurs, as illustrated in the original article of \citet{ohashi2022}. We recall that the non-deformation assumption requires $Ca \ll \epsilon ^2$ which may rewritten using \eqref{eq:h_0dot}

\begin{equation}
Ca_{\dot{a}} \ll \epsilon^2\log(\epsilon).
\end{equation}
In the case where $\epsilon = 0.2$, it follows that $\epsilon^2|\log(\epsilon)|$ yields an approximate value of 0.064 which is indeed much larger than the minimum capillary number examined in the study conducted by \citet{ohashi2022}. 
As a consequence of deformation and the departure of our solution from the bi-spherical solution due 
to the increased film thickness the agreement between formula \eqref{eq:h_0dot} and the experimental data is only noticeable within a relatively limited range of film thickness ($0.25 \lesssim \epsilon \lesssim 0.5$) (see Figure \ref{fig:ohashi}).






\section{Discussion}
\label{sec:conc}

In this article, we have computed the lubrication force between two translating bubbles with time-dependent radii in viscous-dominated flows. Our findings demonstrate that the lubrication theory successfully captures the dominant term in the force expression for small inter-bubble gaps. However, for more accurate results, including $\mathcal{O}(1)$ terms in the force expression is necessary. Subsequently, we applied our findings to two problems in viscous-dominated flows: the coalescence of a small bubble at a free surface and the dynamics of two force-free expanding bubbles. The theory exhibited reasonable agreement with experimental observations within the prescribed range of applicability.

\begin{figure}
\centering
\includegraphics[scale=0.4]{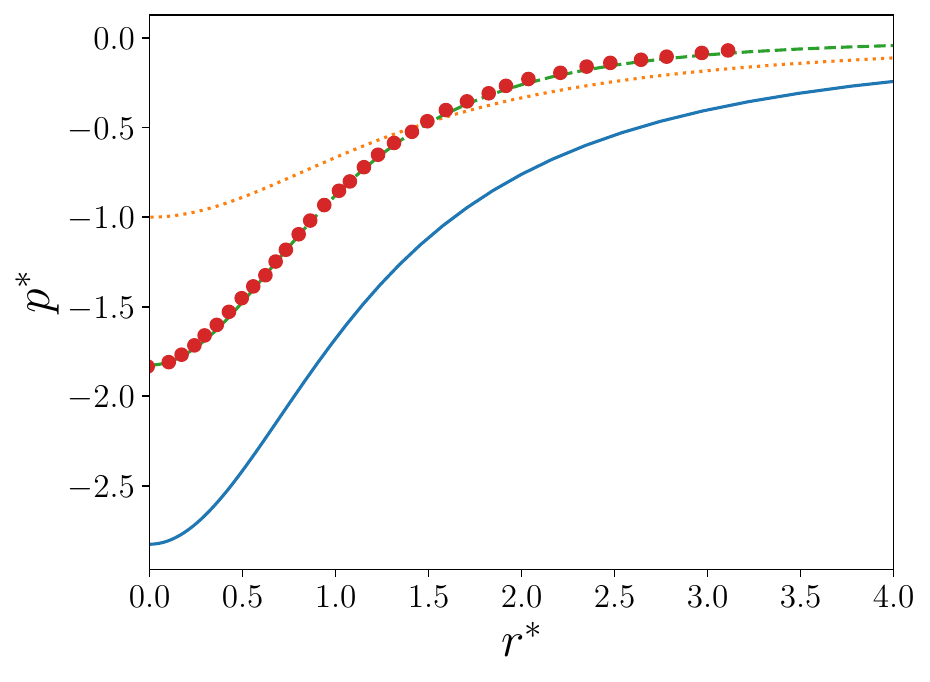}
\caption{Dimensionless pressure within the thin film as a function of the radial distance. \color{blue_py}---\color{black}: formula \eqref{eq:p_lambda} for $\lambda = \epsilon$, \color{green_py} - - \color{black}: $p^* = 2 \frac{\lambda}{\epsilon} \int _{r^*} ^\infty\frac{f^*}{h^*}dr^*$, \color{orange_py} $\cdot \cdot$ \color{black}: formula \eqref{eq:pressure},\color{red_py} $\bullet$ \color{black}: \citet{davis1989} numerical results.}
\label{fig:pressure_lambda}
\end{figure}

In Section \ref{sec:vaka}, it was established that the negligible deformations condition necessitates the Bond number to be very small. In practical applications, it is observed that for a given inclusion size, the Bond number is approximately two orders of magnitude smaller for droplets ascending in liquids than bubbles \citep{balla2020}. Consequently, our theoretical framework may encompass a broader range of droplet sizes than bubble sizes. However, given that droplets typically exhibit higher viscosity ratios than bubbles, there exists a practical need to relax the shear-free boundary condition ($\lambda \ll \epsilon$) and consider scenarios where $\lambda \sim \epsilon$. Although the derivation of the governing equations remains the same (see section \ref{sec:lubrication}), it becomes necessary to incorporate the internal flow within the droplet by accounting for the tangential stress at the interface (as expressed in Equation \eqref{eq:qdm_shear}). Within the Stokes flow regime, the velocity distribution inside the bubble or droplet can be determined by employing the boundary integral formulation of the Stokes equations. \citet{davis1989} have previously obtained an equation characterizing the shear stress distribution along an axisymmetric plane interface as a function of the velocity. Their expression is as follows

\begin{equation}
f^*(r^*)=4\int_0^\infty \phi\left(\frac{R^*}{r^*}\right)\left(\frac{u^*}{R^{*2}}-\frac{1}{R^*}\frac{\partial u^*}{\partial R^*}-\frac{\partial^2 u^*}{\partial R^{*2}}\right)dR^*
\end{equation}
where $\phi$ is a Green function defined in Appendix \ref{app:f}. Although this formulation has been used by various authors \citep{davis1989,rother1997,nemer2013}, the computation of $f^*$ via numerical integration can pose challenges, as evidenced by the recent investigation by \citet{ozan2019}. The numerical integration process is detailed in Appendix \ref{app:f} and a Python script performing the integration can be found here \citep{piersongit}. Given that $f^*$ remains unaffected by the drop radius within the scope of the approximation considered here (nearly flat fluid interfaces), we may substitute $2f^*$ for $f_2^* - f_1^*$ in Equation \eqref{eq:qdm_shear}. Upon integrating Equation \eqref{eq:qdm_shear} with respect to $r^*$ and using Equation \eqref{eq:u}, the resulting pressure profile can be expressed as

\begin{equation}
p^* = -\frac{1}{h^*} + 2 \frac{\lambda}{\epsilon} \int _{r^*} ^\infty\frac{f^*}{h^*}dr^*
\label{eq:p_lambda}
\end{equation}
Numerically integrating the last term in the previous expression produces the function $p^*$ as depicted in Figure \ref{fig:pressure_lambda}. The contribution arising from the non-negligible shear at the interface significantly affects the overall pressure distribution within the film.

The force acting on the bubble (or droplet) can be calculated using the methodology detailed in Appendix \ref{app:f}, resulting in

\begin{equation}
F = -4\pi\mu \bar{a} \dot{d} \left( \log \epsilon - C \frac{\lambda}{\epsilon} \right) .
\label{eq:force_lambda_num}
\end{equation}
where $C \approx 1.3085$. Equation \eqref{eq:force_lambda_num} reveals that the force comprises two distinct contributions. One arises from the pressure within the film when the bubble interfaces experience negligible shear, while the other arises due to the assumption of non-negligible shear at the interface. Both contributions exhibit similar orders of magnitude when $\lambda \sim \epsilon \log \epsilon$. Furthermore, it is evident from \eqref{eq:force_lambda_num} that for small film thickness, i.e., when $\epsilon \log \epsilon \ll \lambda  $, the term associated with non-negligible shear becomes dominant. Additionally, this term is similar to the analysis conducted by \citet{davis1989}. 

Although equation \eqref{eq:force_lambda_num} may be used in practical scenarios involving bubbles or droplets with varying and time-dependent radii, section \ref{sec:viscf} has demonstrated that incorporating the subdominant $\mathcal{O}(1)$ terms in the expansion enhances the accuracy of the results. Given that in the limit where $\lambda \sim \epsilon$, the correction to the $\mathcal{O}(1)$ term derived in part \ref{sec:viscf} is negligible the force on two equal bubbles approaching with a velocity $V$ may be expressed as 




\begin{equation}
F = 8\pi \mu \bar{a} V \left( \log \epsilon - C \frac{\lambda}{\epsilon} + A \right).
\label{eq:force_lambda_num_equals}
\end{equation}
\begin{figure}
\centering
\includegraphics[width=5cm]{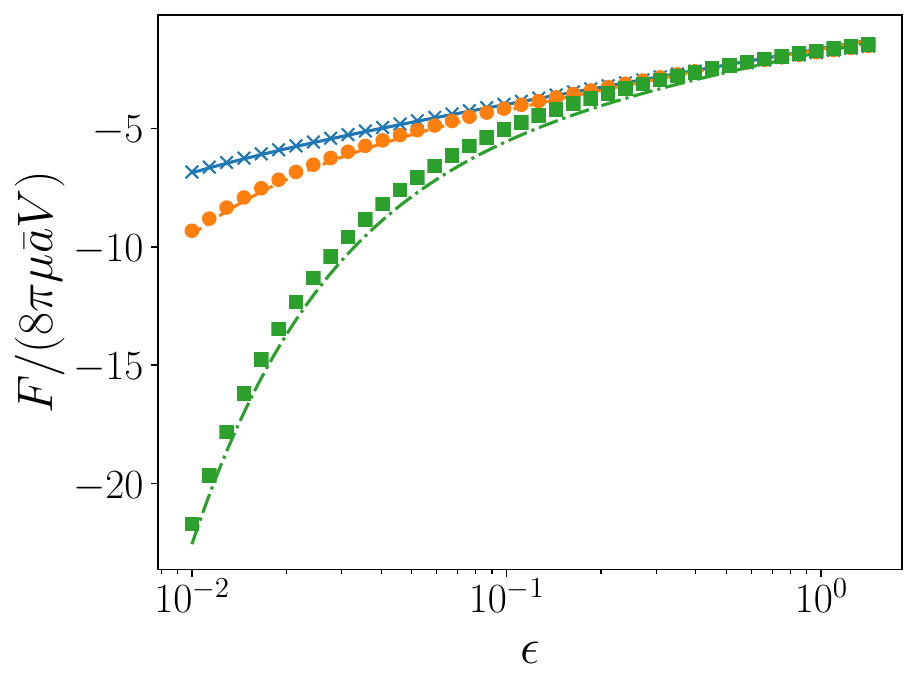}
\caption{Force on two bubbles or drops approaching each other: comparison of formula \eqref{eq:force_lambda_num_equals}, depicted by lines, with bi-spherical coordinate solution (Equation \eqref{eq:bispherical_trans}), represented by symbols. $\lambda = 0.005$ : \color{blue_py} $\times$ \color{black} or \color{blue_py} --- \color{black}, $\lambda =0.025$: \color{orange_py} $\bullet$ \color{black} or \color{orange_py} - - \color{black}, $\lambda =0.125$: \color{green_py} $\blacksquare$ \color{black} or \color{green_py} $- \cdot -$ \color{black}.}
\label{fig:two_drops}
\end{figure}
In figure \ref{fig:two_drops}, we compare this asymptotic solution with the exact solution obtained using bi-spherical coordinates. A very good match is observed between both solutions. Also, one may see that for the highest viscosity ratio considered in the figure, the influence of the term due to non-negligible shear becomes predominant for small $\epsilon$.

Significant effects of gas viscosity are anticipated, particularly in an air-water system where the viscosity ratio ($\lambda \approx 0.018$) is non-negligible. Given the practical importance of the air-water system, it warrants discussion. We anticipate no significant influence from the viscosity ratio prior to film rupture for very small bubbles, typically smaller than $100 \mu m$. However, as illustrated in Fig. \ref{fig:two_drops}, for larger bubbles, the effect of the viscosity ratio becomes non-negligible when the bubbles experience minimal deformation and most probably subsequently when substantial deformation occurs, leading to film drainage \citep{liu2019}. Moreover, since the pressure within the film depends on the viscosity ratio, one may expect a significant influence of the viscosity ratio on the thickness for which bubble deformations occur. The intricate interplay between bubble deformations and viscosity ratio remains a subject for future investigation. This problem could be addressed by employing an asymptotic expansion for small $Ca$ as demonstrated in \citet{yiantsios1990}, for instance. Nonetheless, given that this method is constrained to small deformations, employing numerical simulations may be better suited for investigating scenarios involving significant deformations.


Although we derived the lubrication force in the Stokes limit, one may anticipate obtaining it for moderate Reynolds numbers. Indeed one may consider inertia in the derivation of lubrication equations \citep{chesters1982,howell1996,savva2009} and consider its effect on the lubrication force. We may also compute the subdominant $\mathcal{O}(1)$ terms to improve the accuracy of the solution. One potential approach is to employ direct numerical simulations, as demonstrated in the works of \citet{teng2022,terrington2023}. They explicitly compute the $\mathcal{O}(1)$ term in the force expression in the case of a rotating and possibly translating cylinder nearby to a plane wall. Another related methodology was proposed by \citet{kropinski1995} in their investigation of low Reynolds number flows past a cylindrical body. 
Finally, a natural extension of the present work would involve deriving the theory in potential flow regime \citep{van2002}.


\backsection[Acknowledgements]{The support of V\'eronique Lachet and Benoit Cr\'eton from IFP Energies Nouvelles is gratefully acknowledged. The author is indebted to Professor Masatoshi Ohashi for providing the experimental datas of \citet{ohashi2022}. The author wishes to thank the referees for their valuable comments which helped him to improve the presentation.}

\backsection[Funding]{The financial support of IFP Energies Nouvelles is acknowledged.}

\backsection[Declaration of interests]{The authors report no conflict of interest.}

\appendix

\section{Details on the calculation of the lubrication force}
\subsection{Pressure distribution}
\label{app:pressure}

The viscous term may be separated into two parts. Using equation \eqref{eq:u} the integral of the first part can be expressed as follows
\begin{equation}
2 \int \frac{\partial}{\partial r^*}\left(\frac{1}{r^*}\frac{\partial}{\partial r^*}(r^* u^*)\right) dr^* = \frac{2}{r^*}\frac{\partial}{\partial r^*}(r^* u^*) + C = -\frac{2}{h^{*2}} + C,
\end{equation}
while the integral of the second part reads
\begin{equation}
2 \int \frac{1}{h^*}\frac{\partial h^*}{\partial r^* }  \left(2\frac{\partial u^*}{\partial r^*}+\frac{u^*}{r^*} \right)dr^* = \frac{2}{h^{*2}} - \frac{1}{h^*}+ C
\end{equation}
where $C$ is a constant.
Consequently the antiderivative of the viscous terms is $-1/h^* +C$.

\subsection{Lubrication force}
\label{app:force}
The expression of the lubrication force is given by \eqref{eq:forcen2}. In this appendix, we provide the integral of each term appearing in the lubrification force.
Since,
\begin{equation}
\int _0^{R_\infty^*/\epsilon} \frac{r^*}{h^*}dr^* = [\log(2+r^{*2})]_{0}^{R_\infty^*/\epsilon}
\end{equation}
Then,
\begin{align}
\int _0^{R_\infty^*/\epsilon} \frac{r^*}{h^*}dr^* &= \log\left(2+\frac{R_\infty^{*2}}{\epsilon^2}\right) -  \log(2) \\
                                          &\sim  -2\log \epsilon \quad \text{as } \epsilon \rightarrow 0.
\end{align}
Moreover,
\begin{equation}
\int _0^{R_\infty^*/\epsilon} \frac{r^*}{h^{*2}}dr^* = \left[-\frac{2}{r^{*2}+2}\right]_{0}^{R_\infty^*/\epsilon},
\end{equation}
Therefore
\begin{align}
\int _0^{R_\infty^*/\epsilon} \frac{r^*}{h^{*2}}dr^* &= -\frac{2}{R_\infty^{*2}/\epsilon^2+2} +2 \\
                                            &\sim 2  \quad \text{as } \epsilon \rightarrow 0.
\end{align}



\section{Bi-spherical coordinate solution}
\subsection{Two identical droplets with equal speed} 
\label{app:lub_stokes}
The lubrication force between two non-deformable drops moving with equal speed toward each other has been obtained in closed form by \citet{haber1973} (see also \citep{kim1991}). 
The force may be expressed as 
\begin{equation}
F = 12 \pi \Lambda \mu a V
\label{eq:bispherical_trans}
\end{equation}
where $V$ is the velocity of each drop and $a$ the reduced radius. The coefficient $\Lambda$ reads

\begin{equation}
\Lambda = \frac{2}{3} \sinh \alpha \sum_{n=1}^{\infty} \frac{n(n+1)}{(2n-1)(2n+3)}\left(\frac{A_n(\alpha)+\lambda B_n(\alpha)}{C_n(\alpha)+\lambda D_n(\alpha)}\right)
\end{equation}
where 
\begin{equation}
A_n(\alpha) = 2((n+1)\sinh 2\alpha+2\cosh 2\alpha-2e^{-(2n+1)\alpha)},
\end{equation}
\begin{equation}
B_n(\alpha)=(2n+1)^2\cosh 2 \alpha-2(2n+1)\sinh 2\alpha -(2n+3)(2n-1)+4e^{-(2n+1)\alpha}, 
\end{equation}
\begin{equation}
C_n(\alpha) = 4 \sinh [(n-1/2)\alpha] \sinh [(n+3/2)\alpha],
\end{equation}
\begin{equation}
D_n(\alpha) = 2 \sinh[(2n+1)\alpha] - (2n+1)\sinh 2\alpha
\end{equation}
The parameters $\alpha$ is related to the film thickness as  $1 + \epsilon ^2/4 = \cosh \alpha $. 

\subsection{Two identical bubbles with time-dependent radii}
\label{app:visc}

In this appendix, we calculate the force exerted on two identical bubbles with time-dependent radii using the bi-spherical solution proposed by \citep{michelin2018}. Due to the identical nature of the bubbles, the plane of symmetry that separates them is a shear-free boundary. Consequently, the coefficients $A_n$ and $C_n$ defined in \citep{michelin2018} are both zero, leading to the following expression for the force

\begin{equation}
F = \frac{8\pi \sqrt{2}\mu \dot{a} a^2}{k}\sum_{n=1}^{\infty}\left(\frac{2n+1}{4n+2}\right)(B_n+D_n)
\label{eq:bispherical} 
\end{equation}
where $\dot{a}$ is the time rate of change of the bubble radius and
\begin{equation}
B_n = \frac{-U_n''+(n+3/2)^2U_n}{\sinh[(n-1/2)\eta]} \quad \text{and} \quad D_n = \frac{U_n''-(n-1/2)^2U_n}{\sinh[(n+3/2)\eta]}.
\end{equation}
The functions $U_n$ and $U_n''$ are defined as
\begin{align}
U_n(\eta) = &-\frac{3\sqrt{2}\sinh^2(\eta/2)}{2(2n+1)}  \left(\frac{e^{-(n+3/2)\eta}}{2n+3} - \frac{e^{-(n-1/2)\eta}}{2n-1}\right) \\
            &- \frac{\delta_{n1}\sqrt{2}}{3}(e^{\eta/2}-e^{-\eta/2})+\frac{\sqrt{2}\sinh^2\eta}{2(2n+1)}e^{-(n+1/2)\eta},
\end{align}
and
\begin{align}
U_n''(\eta) =& -\frac{3\sqrt{2}\sinh \eta}{8}  \left(-\frac{2\sinh^2 \eta e^{-(n+1/2)\eta}}{1+\cosh\eta}+(2n+3)e^{-(n-1/2)\eta}-(2n-1)e^{-(n+3/2)\eta}\right) \nonumber\\
 &- \frac{\delta_{n1}}{\sqrt{2}}(e^{\eta/2}-e^{-\eta/2}) - \left(n-\frac{1}{2}\right)\left(n+\frac{3}{2}\right)U_n(\eta),
\end{align}
where $k = \sqrt{(h_0+2a)^2/4-a^2}$ and $\sinh \eta = k/a$. In practice we truncate the infinite sum to a finite number $N \geq 1000$.

\section{Computation of the tangential shear stress $f^*(r^*)$ and the force $F^*$}
\label{app:f}
In this appendix, we give details on the way $f$ and $F$ are computed. The function $\phi$ can be expressed as follows \citep{davis1989,rother1997,nemer2013}


\begin{equation}
\phi(x) = \frac{1}{2\pi}\left[\frac{1+x^2}{1+x}K\left(\frac{4x}{1+x^2}\right)-(1+x)E\left(\frac{4x}{1+x^2}\right)\right]
\end{equation}
where $x = R/r$ and $K$ and $E$ are the first and second-kind elliptic integrals. The primary difficulty encountered in the computation of function $f$ arises from the divergence of $\phi$ as the radius $R^*$ approaches $r^*$. Indeed,


\begin{equation}
\phi\left(\frac{R^*}{r^*}\right) \sim -\frac{1}{2\pi}\ln\left(\frac{| R^*-r^* |}{r^*}\right) \quad \text{as} \quad  R^* \rightarrow r^*. 
\end{equation}
To perform the integration we define two integrals by adding or substrating the previous "weak" logarithm singularity


\begin{equation}
f_i^*(r^*)=4\int_0^\infty \left[\phi\left(\frac{R^*}{r^*}\right)+\frac{1}{2\pi}\ln\left(\frac{| R^*-r^* |}{r^*}\right)\right]\left(\frac{u^*}{R^{*2}}-\frac{1}{R^*}\frac{\partial u^*}{\partial R^*}-\frac{\partial^2 u^*}{\partial R^{*2}}\right)dR^*,
\end{equation}

\begin{equation}
f_s^*(r^*)=4\int_0^\infty -\frac{1}{2\pi}\ln\left(\frac{| R^*-r^* |}{r^*}\right)\left(\frac{u}{R^{*2}}-\frac{1}{R^*}\frac{\partial u^*}{\partial R^*}-\frac{\partial^2 u^*}{\partial R^{*2}}\right)dR^*,
\end{equation}
The first integral is calculated numerically whereas the second integral can be obtained analytically, resulting in 
\begin{equation}
f_s^*(r^*)= -\frac{2+4\ln r^*+3\pi r^*+2r^{*2}+\pi r^{*3}}{\pi (1+r^{*2})^2}.
\end{equation}
In order to compute the pressure within the thin film, as expressed by formula \eqref{eq:p_lambda}, we decompose it into two distinct contributions. The first contribution, denoted as $p^*_{\lambda \ll \epsilon}$, arises solely from the viscous stress within the film when the shear at the interface is negligible, and can be expressed as $p^*_{\lambda \ll \epsilon} = -1 /h^*$. The second contribution, denoted as $p^*_{\lambda \sim \epsilon}$, emerges due to the non-negligible shear at the interfaces and is given by the expression $p^*_{\lambda \sim \epsilon} = 2 \frac{\lambda}{\epsilon} \int _{r^*} ^\infty\frac{f^*}{h^*}dr^*$. By noting that  $f_i(r^*) -4/\pi\ln r^*$ and $f_s+4/\pi \ln r^*$ can be integrated numerically we can compute $p^*_{\lambda \sim \epsilon}$ as depicted in Figure \ref{fig:pressure_lambda}. 


We now compute the contribution fo the force due to the non-negligible shear at the interfaces denoted as $F^*_{\lambda\sim \epsilon }$. This contribution may be expressed as follows

\begin{equation}
F^*_{\lambda\sim \epsilon } = 2 \pi \int _0 ^{R_\infty ^* /\epsilon}  p^*_{\lambda \sim \epsilon} dr^*. 
\end{equation}
This last expression can be simplified in several ways. First, the upper limit of the preceding expression should be in the range $1 \ll R_\infty ^* /\epsilon \ll 1/\epsilon$. However, given that the parameter $p^*$ tends towards zero as the radial coordinate $r^*$ approaches infinity, we may substitute the upper limit with $r^*=\infty$ \citep{davis1989}. Second, this integral may be integrated by parts. Assuming that $p^*r^{*2}$ tends towards zero as $r^*$ approaches infinity, we get 

\begin{equation}
F^*_{\lambda\sim \epsilon } = 2 \pi \frac{\lambda}{\epsilon} \int _0 ^{\infty} \frac{f^*}{h^*} r^{*2} dr^*.
\end{equation}
Performing the numerical integration yields an approximate result of 

\begin{equation}
F^*_{\lambda\sim \epsilon } \approx -16.4435605 \frac{\lambda}{\epsilon}. 
\end{equation}
This approximated result can be compared to the exact coefficient as reported by \citet{kim1991} $-\frac{3 \sqrt{2}}{8}\pi^3 \approx -16.4435614$. For implementation details, a Python script facilitating the integration process can be found here \citep{piersongit}.

\bibliography{biblio.bib}
\bibliographystyle{apalike}

\end{document}